\documentstyle[12pt]{article}
\begin{document}
\title{Nonlinearity Effect on 1D Periodic and Disordered Lattices } 
\author{K. Senouci$^{1}$, N. Zekri $^{1}$, H. Bahlouli$^{2}$ and A.K. Sen $^{3}$\\
AS-ICTP, 34100 Trieste Italy\\
$^{1}$Laboratoire d'Etude Physique des Materiaux, Departement de\\
Physique, UST Oran, B. P. 1505, Oran El-M'naouer, Algerie. \\
$^{2}$Physics Department, KFUPM, Dhahran 31261, Saudi-Arabia. \\
$^{3}$LTP Division, Saha Institute for Nuclear Physics, \\
1/AF Bidhannagar, Calcutta 700 064, India }
\maketitle
\begin{abstract}
The Kronig-Penney model is used to study the effect of nonlinear interaction
on the transmissive properties of both ordered and disordered chains. In the
ordered case, the nonlinearity can either localize or delocalize the electronic
states depending on both its sign and strength but there is
a critical strength above which all states are localized. In the 
disordered case, however, we found that the transmission decays as $T\sim
L^{-\gamma}$ around the band edge of the corresponding periodic system.
The exponent $\gamma$ is independent of the strength of the nonlinearity in
the case of disordered barrier potentials, while it varies with this strength
for mixed potentials.

\end{abstract}

Keyword: Band spectrum, non-linearity, interaction, disorder.
\newpage

\section{Introduction}

\hspace{0.33in}
Wave propagation in nonlinear media is a subject of recent intensive
research \cite{Review}. The study of this phenomenon is of great  
practical importance in the understanding of transport properties of 
superlattices \cite{Super}, electronic behavior in mesoscopic devices and
optical phenomena in general. The nonlinear Schr\"odinger equation has been 
studied extensively in recent years and served as a prototype for studying 
nonlinear phenomena. The origin of the nonlinearity in the Schr\"odinger 
equation might correspond to different physical phenomena. In electronic systems 
it would correspond to Coulomb interaction between confined electrons 
while in a superfluid it corresponds to the Gross-Pitaevsky equation which 
attracted much interest in recent years in the area of Bose-Einsten 
condensation of trapped bosonic atoms \cite{BEC}. One then uses the usual 
technique, as in linear systems, to deduce transmission and related 
properties of interest. However, there are differences though with the linear 
problem. Most important for us is the fact that the transmission is not 
uniquely defined. In contrast to the linear case it is no longer equivalent 
to study the transmission for a fixed input (normalized incident wave) or 
fixed output (normalized transmitted wave). This non-equivalence originates 
from the fact that for a given output, there is one and only one solution to 
the given problem. In contrast, for a fixed input, there is at least one 
solution to the problem but, because of the nonlinearity, there might be
more than one solution for a given system length \cite{Devil}. In particular,
it is believed that this non-uniqueness gives rise to multistability and noise
and might originate a chaotic behavior in certain systems \cite{Wan,Souk}. 

\hspace{0.33in} From the theoretical point of view we expect 
new effects to arise due to the competition between the well known localizing 
effects of the disorder and the delocalizing effect due to the nonlinear 
interaction in an appropriate regime. Anderson's theory predicts that the wave
function of a non-interacting electron moving in a one dimensional lattice
with on-site energetic disorder is localized even for an infinitesimal amount
of disorder  \cite{Ander}. Thus in the linear regime but in the presence of
disorder, for a given incident wave with wavenumber $k$ (or an electron with
energy $E$), the transmission 
coefficient decays exponentially with the system length. On the other hand, 
the decay of the transmission is much slower in nonlinear systems. Actually 
a power-law decay of the transmission was already obtained in nonlinear 
systems with on-site disorder \cite{Devil,Abdul}. However, Kivshar et al. 
\cite{Kiv}, while studying the propagation of an envelope soliton in a  
1D disordered system, have found that the decay is actually not a power-law
type and that strong nonlinearity washes out localization effects. This means
that above a certain critical value of the nonlinearity strength we can have
wave propagation in nonlinear disordered media, which is a situation of 
great practical interest. Molina et al.\cite{Molina}, on the other hand, 
studied the transport properties of a nonlinear disordered binary alloy
using a tight binding Hamiltonian.  They have confirmed the power law
behavior of the transmission but concluded that the decay exponent does not
depend on the degree of nonlinearity and that delocalization 
disappears for large nonlinearities. 

\hspace{0.33 in} It is the purpose of this work to study 
how the decay of the transmission is affected by nonlinear interactions in 
general for disordered systems and its effect in periodic systems. In 
particular, in a recent work on nonlinearity effect on periodic systems 
\cite{Zekri}, we found that the bandwidth decreases when the lattice
potential has the same sign as the 
nonlinear interaction coefficient while in the case of opposite signs the 
bandwidth increases and some states appear in the bandgap of the 
corresponding linear periodic system. We study here the scaling properties of the 
transmission at these gap-states in order to know how the nature of 
the eigenstates behave with nonlinearity. 

\section{Model description}

\hspace{0.33in} Due to the above mentioned non-uniqueness problem 
we will restrict ourselves 
to a uniquely defined situation where the output is fixed and one is 
interested in finding the necessary input. Leaving this issue aside 
we would like to investigate the effect of nonlinearity on the transmission 
of an ordered and disordered Kronig-Penney lattice model. We use the 
following standard model to describe this system \cite{Cota}

\begin{equation}
\left\{ -\frac{ d^{2} }{ dx^{2} } + \sum_{n}  ( \beta_n 
 +  \alpha \left| \Psi (x) \right| ^{2} ) 
 \delta(x-n) \right\} \}\Psi (x) = E\Psi (x) 
\end{equation}

\noindent Here $\Psi (x)$ is the single particle wavefunction at $x$,
$\beta_n $ the potential strength of the $n-th$ site, $\alpha $ 
the nonlinearity strength and $E$ the single particle energy in units of 
$\hbar^{2}/2m$ with $m$ being the electronic effective mass. For simplicity 
the lattice spacing is taken to be unity in all this work. The potential 
strength $\beta_n$ is picked up from a random distribution with 
$ -W/2 < \beta_n < W/2 $ for the mixed potentials case and 
$0 < \beta_n < W$ for the potential barriers case ($W$ being the degree of disorder). 
The local nature of the nonlinear interaction in (1) does not stem only from 
its simplicity in numerical computation, but also from a physical point of view
that many of the interactions leading to nonlinearity are of local nature such
as an on-site Coulomb interaction. From the computational 
point of view it is more useful to consider the discrete version of this 
equation which is called the generalized Poincar\'e  map and can be derived 
without any approximation from equation (1). It reads \cite{Sanchez}

\begin{equation}
\Psi _{n+1} = \left[ 2\cos k + \frac{\sin k}{k}( \beta_n 
+ \alpha |\Psi_{n}|^{2} ) \right] \Psi _{n} - \Psi _{n-1} 
\end{equation}

\noindent where $\Psi _n$ is the value of the wavefunction at site $n$ and 
$k=\sqrt{E}$. This representation relates the values of the wavefunction at
three successive discrete locations along the x-axis without restriction on
the potential shape at those points and is very suitable for numerical
computations. The solution of equation (2) is done iteratively by taking for 
our initial conditions the following values at sites $1$ and $2$ : 
$\Psi _1=$ $\exp (-ik)$ and $\Psi _2=$ $\exp (-2ik)$. We consider here an 
electron having a wave vector $k$ incident at site $N+3$ from the right 
(by taking the chain length $L=N$, i.e. $N+1$ scatterers ). The transmission 
coefficient ($T$) can then be expressed as

\begin{equation}
T = \frac{4\sin^{2}k}{ |\Psi_{N+2} - \Psi_{N+3}\exp (-ik) | ^{2} } 
\end{equation}

\noindent Thus $T$ depends only on the values of the wavefunction at the end 
sites, $\Psi _{N+2}$, $\Psi _{N+3}$ which are evaluated from the iterative 
equation (2). 

\section{Results}

\hspace{0.33 in} First let us examine how the allowed bands and band gaps in 
the periodic systems get affected by the nonlinear interaction. The 
nonlinearity is expected under certain conditions to delocalize the electronic 
wavefunction \cite{Devil,Cota}. Therefore, in the framework of the transmission 
spectrum a decrease of the width of the bandgap will signal delocalization 
while an increase in the bandgap will signal localization effect. To explain 
qualitatively the behavior of the transmission for different signs of the 
nonlinear interaction, we first start with a simple double barrier structure. 
In a recent work we examined the transmission spectrum for this structure but 
we restricted ourselves to small nonlinearity strengths. In Figure 1, we show 
the effect of nonlinearity on the first two resonances of both double barrier 
and double potential well systems. In the case of barriers Fig.1a shows that 
for positive $\alpha$ the resonances get 
displaced to higher energies and become sharper. As we increase $\alpha$ the
valleys deepen which is a signature of confinement within the well between 
the two barriers. For negative $\alpha$, Figure 1b shows that for small values 
$ |\alpha| < \beta$ , the resonances get displaced to lower energies while the 
valleys increase and get more and more suppressed as we increase $\alpha $ 
in magnitude. Thus one can conclude that for small values of $\alpha$ , the 
gap gets suppressed  with increasing values of $\alpha $ provided that 
$|\alpha| < |\beta| $. On the other hand for larger values of the nonlinearity,
$|\alpha| > |\beta| $, the effect is reversed, that is the gap gets larger 
and larger. 

\hspace{0.33 in} If we consider a double potential wells instead (Figs.2) this 
behavior is reversed. Thus for negative nonlinearity (Fig.2a) the valleys 
become deeper while they become more and more suppressed for positive 
nonlinearity as shown in Fig2b and similarly to the case of the barriers the 
valleys start becoming deeper for $|\alpha| > |\beta|$. 
In summary, the nonlinear interaction seems to delocalize the 
electronic states when it is repulsive (attractive) for potential barriers 
(potential wells) and the nonlinearity strength satisfies $|\alpha| < |\beta|$.
For all the other cases it seems to localize the eigenstates. In fact the 
delocalization can be simply explained by the fact that the effective 
potential in (1) tends to vanish. Thus when the on-site potential and the 
interaction potential (represented by nonlinearity) have opposite signs the 
effective potential decreases in Eq.(1) and vanishes for $|\alpha| = |\beta|$. 
Therefore, the electron tends to become free in this case. When the nonlinear 
strength increases the effective potential starts increasing and the electron 
will 'see' the effective potential. 

\hspace{0.33 in} However, we found in the previous work \cite{Zekri} that 
the delocalization (narrowing of the bandgap) in periodic systems appears as 
resonant transmission states (sometimes not overlapping) in the gap. We try 
to examine the nature of these states in the gap of the corresponding linear 
periodic system in the presence of nonlinearity. 
To this end, we choose an energy ($E=11$) in the bandgap of the periodic potential 
barriers and another one ($E=9$) in that of the potential wells. Obviously, in the 
absence of nonlinearity and for finite systems the transmission coefficient decreases 
exponentially with the length scale at these energies (as shown in Figure 3). If we 
switch on the nonlinear interaction (with the sign chosen so as to 
have a delocalization following the above 
discussion) we find that the transmission coefficient (or equivalently the 
wavefunction) becomes Bloch like both for potential barriers (Figure 4a) and potential 
wells (Figure 4b). It is shown in these figures that when the nonlinear strength 
(in absolute value) increases but remains smaller than the absolute value of the 
potential strength ($|\alpha| < |\beta|$) the amplitude of the transmission 
oscillations becomes larger (while its period increases) reaching a constant unity 
transmission at the critical strength ($|\alpha_c|=|\beta|$) while for larger 
nonlinearity strengths ($|\alpha| > |\beta|$) the amplitude of these oscillations 
keeps increasing and its period decreases. This behavior means that the eigenstates 
in the gap region of the corresponding linear systems become extended even for a 
small amount of nonlinearity but the transmission is maximum at the critical nonlinear 
strength $\alpha_c$ (or in other words the resistance vanishes at this critical 
strength). 

\hspace{0.33 in} In order to explain qualitatively this delocalization, we note that 
the nonlinear term in Eq.(1) contains  $|\Psi^2|$ which behaves as the 
inverse of the transmission coefficient from Eq.(3). Thus for decreasing transmission 
this modulus increases while if the transmission 
is close to unity it decreases. Therefore, in the gap region, since the transmission 
coefficient decreases with the length scale, $|\Psi|^2$ increases and consequently 
the effective potential decreases which leads to the increase of the transmission and 
so on. The transmission oscillates then with the length scale and its period depends 
on the speed of the variation of $|\Psi|^2$ which depends on the nonlinearity 
strength. If this strength increases, we reach rapidly the condition of vanishing 
effective potential and the variation of $|\Psi|^2$ is slow (and the period of 
oscillations is large) while for very small strengths this modulus starts increasing 
rapidly up to the condition of vanishing effective potential where it becomes very 
large, and then this effective potential increases rapidly leading to smaller periods 
of the transmission oscillations. We note here that the transmission never decays with 
the length even for high nonlinearity strength. 

\hspace{0.33 in} Let us now examine the effect of disorder on the nonlinear 
Schr\"odinger equation. We consider here two kinds of disorder as discussed above 
(mixed disorder and potential barriers disorder) in order to check the 
kind of disorder dependence of the power law behavior observed in recent 
works \cite{Molina,Cota}. We note here 
that we observe a power-law decay of the transmission near the band edges of 
the corresponding periodic system (i.e., around $k=n \pi /a$, $n$ being a positive 
integer number and the lattice parameter $a$ is taken here to be unity). For all 
other energies, the decay of the transmission with the length becomes either 
exponential or even stronger (we did not show these results here).  In this connection,
we would like to remark that the energy taken by Cota et al. 
\cite{Cota} (their model is exactly the same as our mixed potentials model) is $E=5$ 
instead of $E \simeq 10$ (probably due to a misprint in their paper). As found by 
Cota et al. \cite{Cota}, for $E=5$ the mixed case shows a power-law decay above a critical 
nonlinearity strength (actually they did not fit a power-law behavior for the
strengths of $\alpha = 10^{-15}$ and $10^{-10}$). In contrast, what we find is that
for $E=5$, the transmittance decays exponentially for small disorders and small
$\alpha$, but faster than exponentially for larger disorder and/ or nonlinearity. 
However, if we choose $E=10$ (which is close to the band-edge for a periodic system
but inside the gap), there is a finite size effect and the power-law decay of the
transmission is observed only above a characteristic length $L_{c}$ which seems to
decrease with nonlinearity strength as clearly shown in Fig.5a (below this $L_c$, 
the transmission is 
exponentially decaying). Further, even for very small nonlinearity strengths 
(e.g., $\alpha = 10^{-15}$ and $10^{-10}$) there is a crossover to a power-law 
decay of the transmittance for $L > L_c$. This power-law behavior is also 
shown in the case of disordered barrier potentials (Fig.5b) but the characteristic
length $L_c$ seems 
to be smaller. We did not show here the case of disordered potential wells because it 
is similar to that of the potential barriers but for a positive sign of the 
nonlinearity.  

\hspace{0.33 in} As shown in Figs.5, the exponent of the power law decay 
$\gamma$ seems to be slightly dependent on the nonlinearity strength for the mixed 
case while it seems to be almost constant for disordered barrier potentials. 
This behavior is confirmed in Figure 6 where we fitted the power-law behavior only 
above $L_{c}$. This figure shows a qualitative agreement with the 
results of Cota et al. \cite{Cota} (except for the fact that there is no critical
$\alpha$) for a mixed disorder while for barrier type 
disorders, the exponent is smaller and seems to become independent of the 
nonlinearity strength. This last result has been also found by Molina et al. 
\cite{Molina} for disordered binary alloys who first of all used a tight binding
Hamiltonian and then lumped the disorder in the nonlinear coefficient itself (this
is entirely different from the model we used).  On the other hand, Cota et al.
used the same model as we do, but the behavior observed by them is not universal 
and depends on the kind of disorder. Indeed, for disordered potential barriers the 
negative nonlinearity strength tends to delocalize the eigenstates  as shown for the 
double barriers (Fig.1a) and for the periodic systems (Figs.4) while in the mixed 
case, there is always a competion between the delocalization in potential barriers 
and the strong localization in the remaining potential wells which increases the 
characteristic length $L_{c}$.  We would also like to point out that the power-law
behavior becomes very sensitive on some particular configurations in the large length
scale, and tends to give very large values of the resistance making the 
calculations on the average properties unstable.

\section{Conclusion}
 
\hspace{0.33in} We studied in this paper the effect of nonlinearity both on double 
barriers, periodic and disordered systems using a simple Kronig-Penney Hamiltonian. 
We found in the double barriers system a range of nonlinearity strengths for which 
the delocalization takes place and a critical nonlinearity strength above which 
the behavior is reversed (At this critical value the transparency becomes unity). 
It seems also that the nonlinearity suppresses the gap in periodic systems. Indeed, 
for finite size systems, the transmission for energies corresponding to the gap in 
infinite systems is exponentially decaying while, with any 
small amount of nonlinearity it becomes Bloch like. Finally in the presence of 
disorder and in the regime of nonlinearity strengths delocalizing the gap states of
the periodic system, 
we found that the transmission becomes power-law decaying around the band edges 
of the corresponding periodic system while for other energies the transmission is 
at least exponentially decaying if not faster. The exponents of the power-law behavior
(above the $\alpha$-dependent crossover length scale $L_c$) of the transmittance depends on
the nonlinearity strength for mixed systems in qualitative agreement with 
the results of Cota et al. \cite{Cota}, while it seems to be constant for potential 
barriers in agreement with the results of Molina et al. \cite{Molina} even though the 
system used by these last authors is different from ours (they used a tight binding 
model with a disorder in the nonlinearity strength itself). Therefore, the variation of 
this exponent with nonlinearity depends strongly on the type of disorder used and is 
not universal as claimed recently \cite{Cota}. On the other hand, this exponent is 
much larger in mixed systems than in disordered potential barriers. It is then 
interesting to examine within this model the effect of disordered nonlinearity on 
the transport properties in order to compare them with the results of Molina et al. 
\cite{Molina}. Also, this power-law behavior is observed only
above a characteristic length $L_{c}$. It is then intersting to study the finite 
size effect of this behavior. Furthermore, since metallic 
and insulating behaviors are well characterized by the statistical 
properties of their transport coefficients \cite{Fluct}, it seems to be 
adequate to examine the transition from exponentially localized states 
in linear disordered systems to power law decaying states in nonlinear ones
using the above technique. 
 
\vspace{0.2 in}

{\bf ACKNOWLEDGEMENTS }

We would like to acknowledge the support of ICTP during our stay at Trieste.
H. B. also would like to acknowledge the support of the physics department at King
Fahd University of Petroleum and Minerals ( KFUPM ) during the progress of
this work.

\newpage

\begin{center}
{\bf Figure Captions}
\end{center}

\bigskip

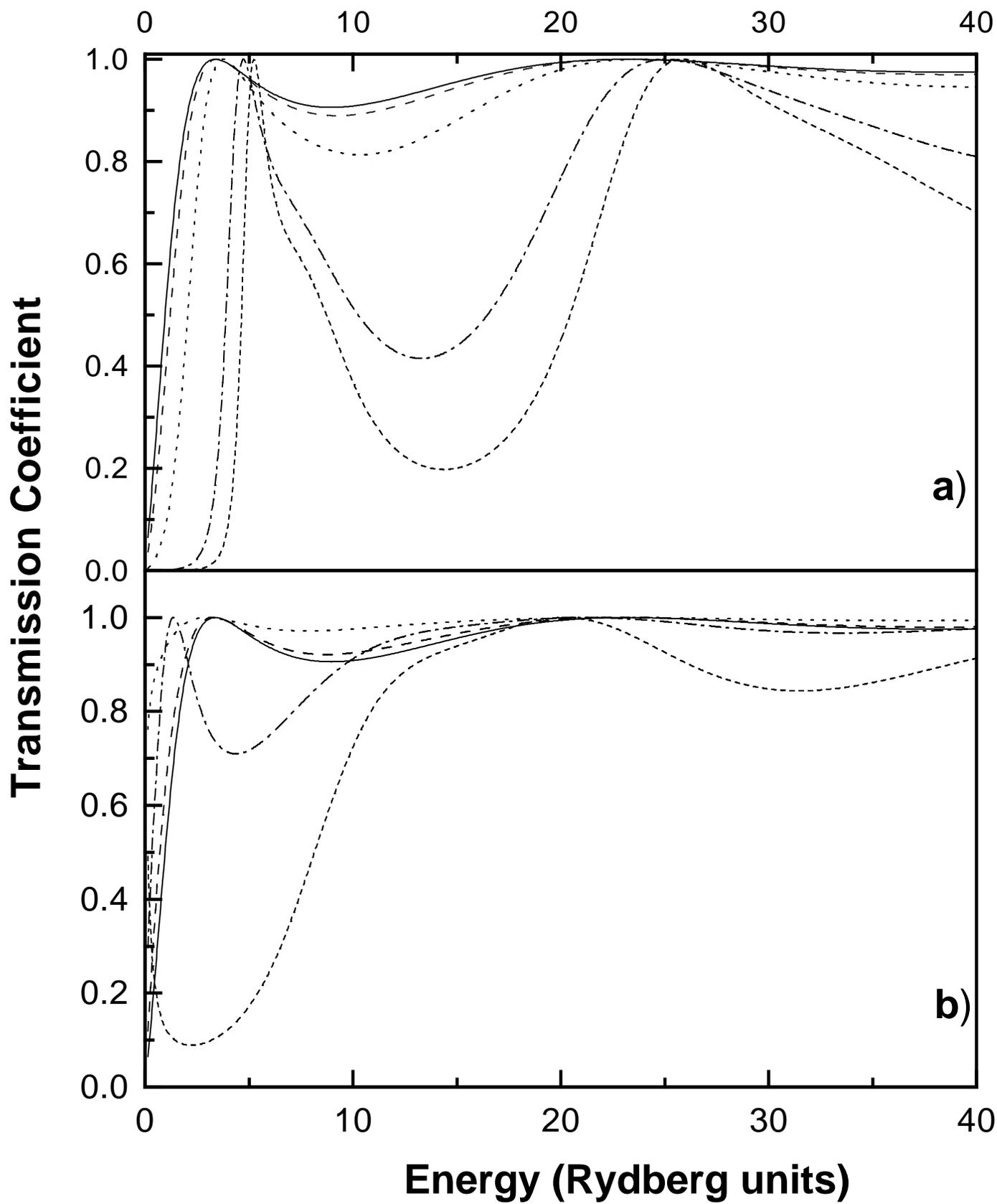
{\bf Fig.1} Transmission coefficient versus energy for a double barrier with 
$\beta = 1 $ , $ |\alpha|$ = $0.$ (solid curve), $0.1$ (dashed curve), $0.5$ 
(dotted curve), $2$ (dash-dotted curve) and $3$ (short dashed curve). a) $\alpha > 0$, 
b) $\alpha < 0$. 

\bigskip

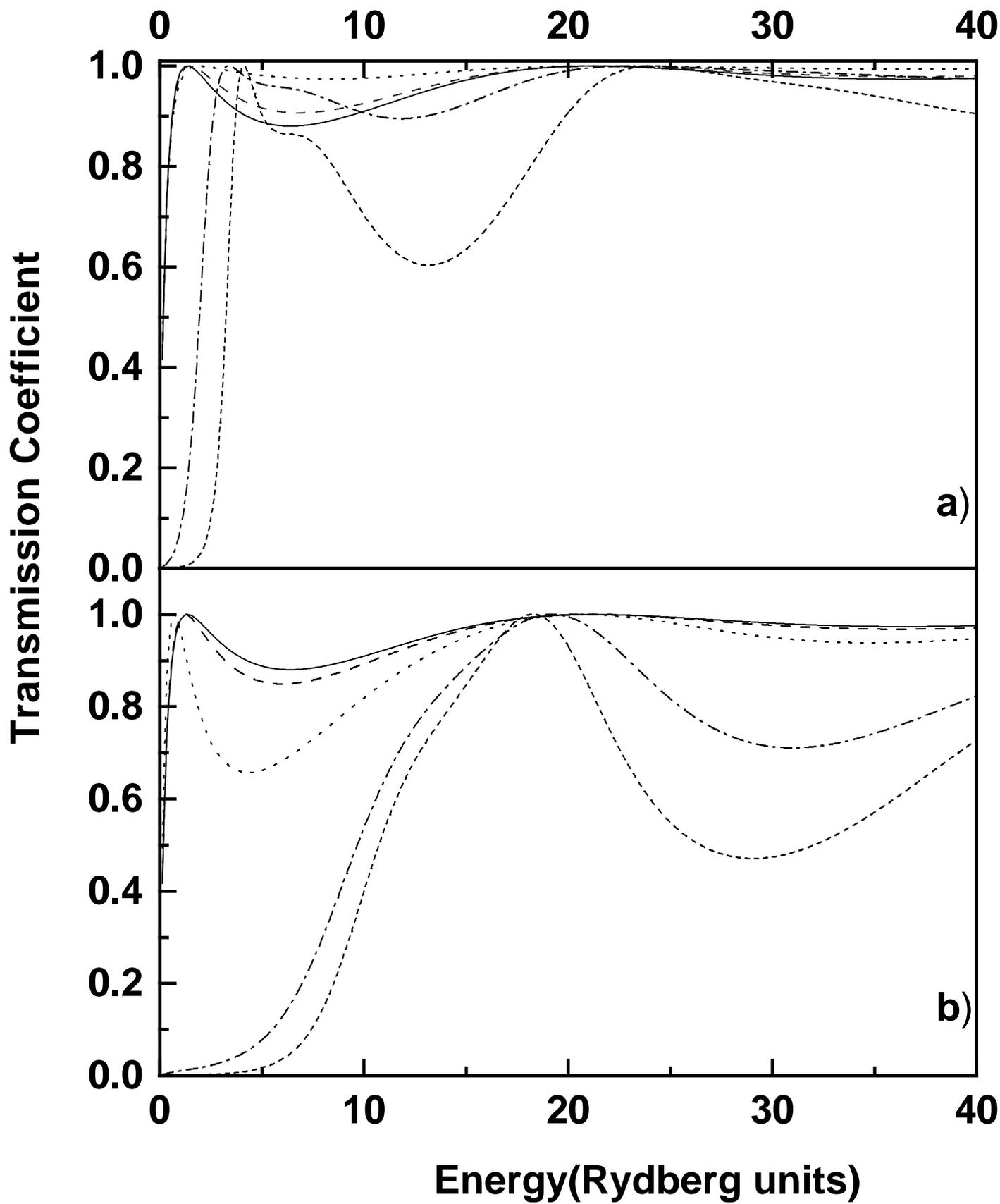
{\bf Fig.2 } Same as Fig.1 for double well ($\beta = -1 $). 

\bigskip

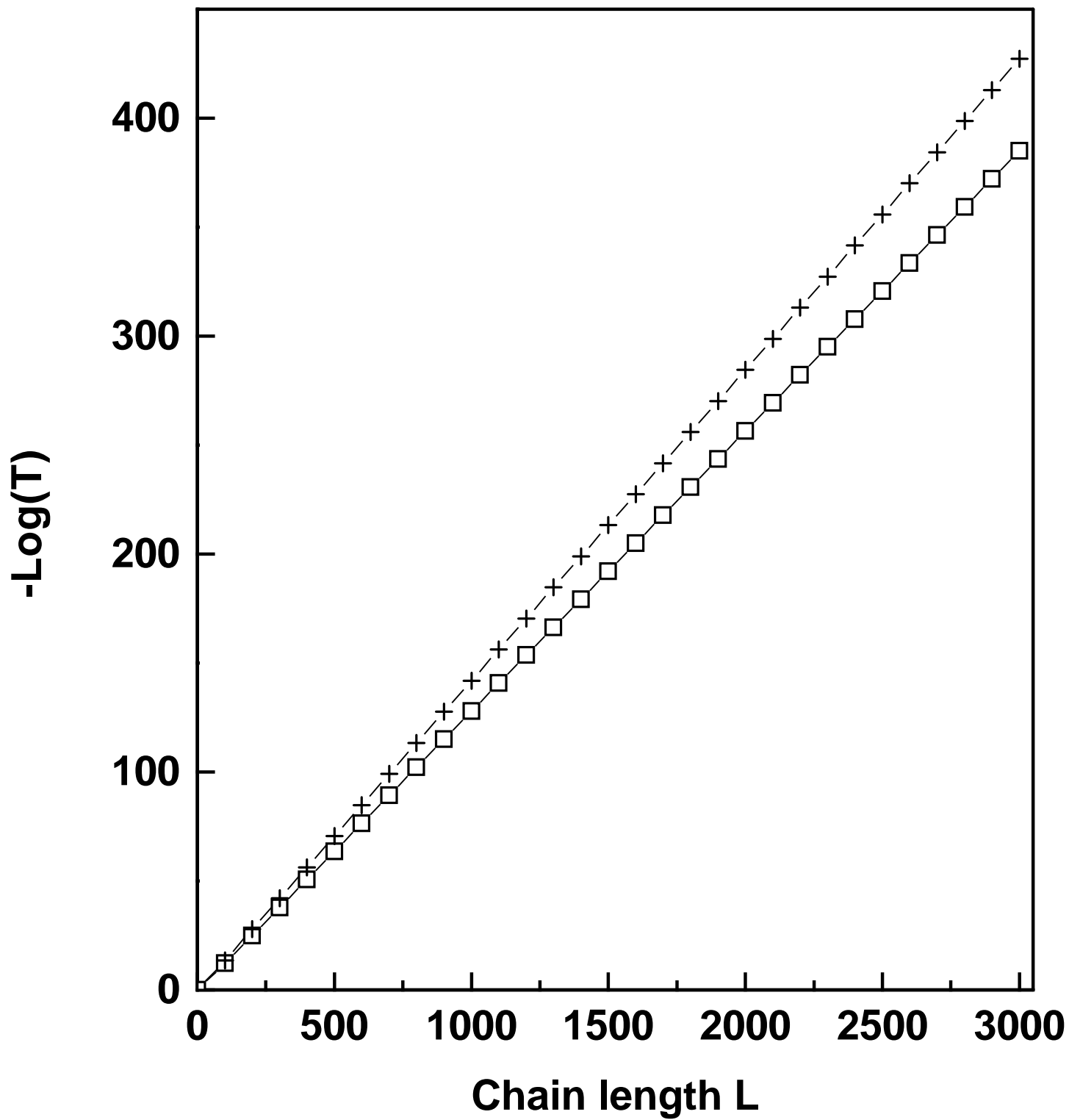
{\bf Fig.3 } $-LogT$ versus $L$ for linear periodic system ($\alpha = 0.$) for both
potential barriers $\beta=1,E=11$ (open squares) and potential wells $\beta=-1,E=9$ 
(cross symbols +).

\bigskip

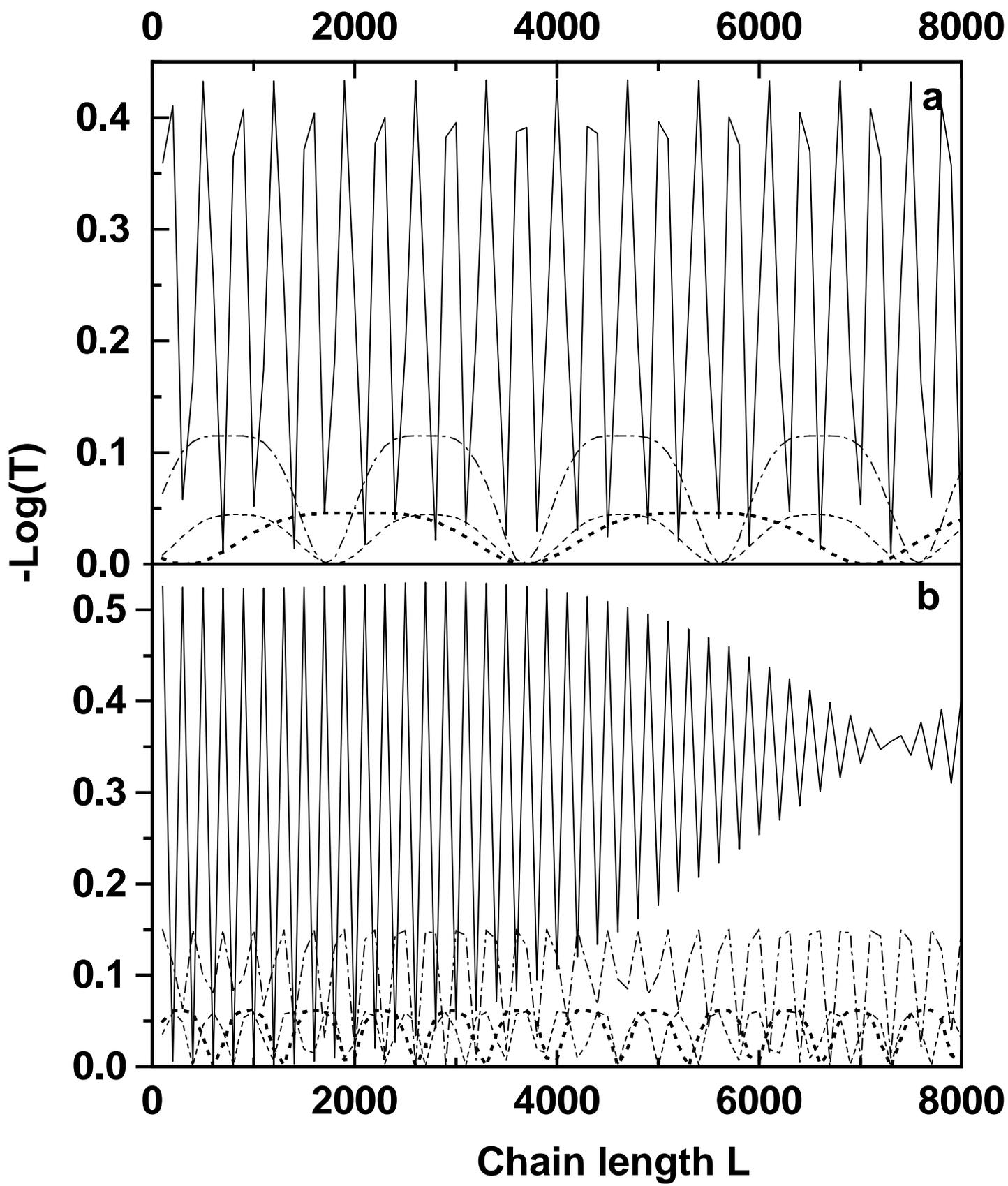
{\bf Fig.4 } $-LogT$ versus $L$ for $|\alpha| $= $0.1$ (solid curve) , $0.5$ (dashed 
curve), $2.$ (thick dotted curve) and $3.$ (dash-dotted curve) for a) potential barriers 
($\beta$ = $1$, $E = 11$  and $\alpha < 0 $) and b) potential wells ($\beta$ = $-1$, 
$E = 9$ and $\alpha < 0 $).

\bigskip

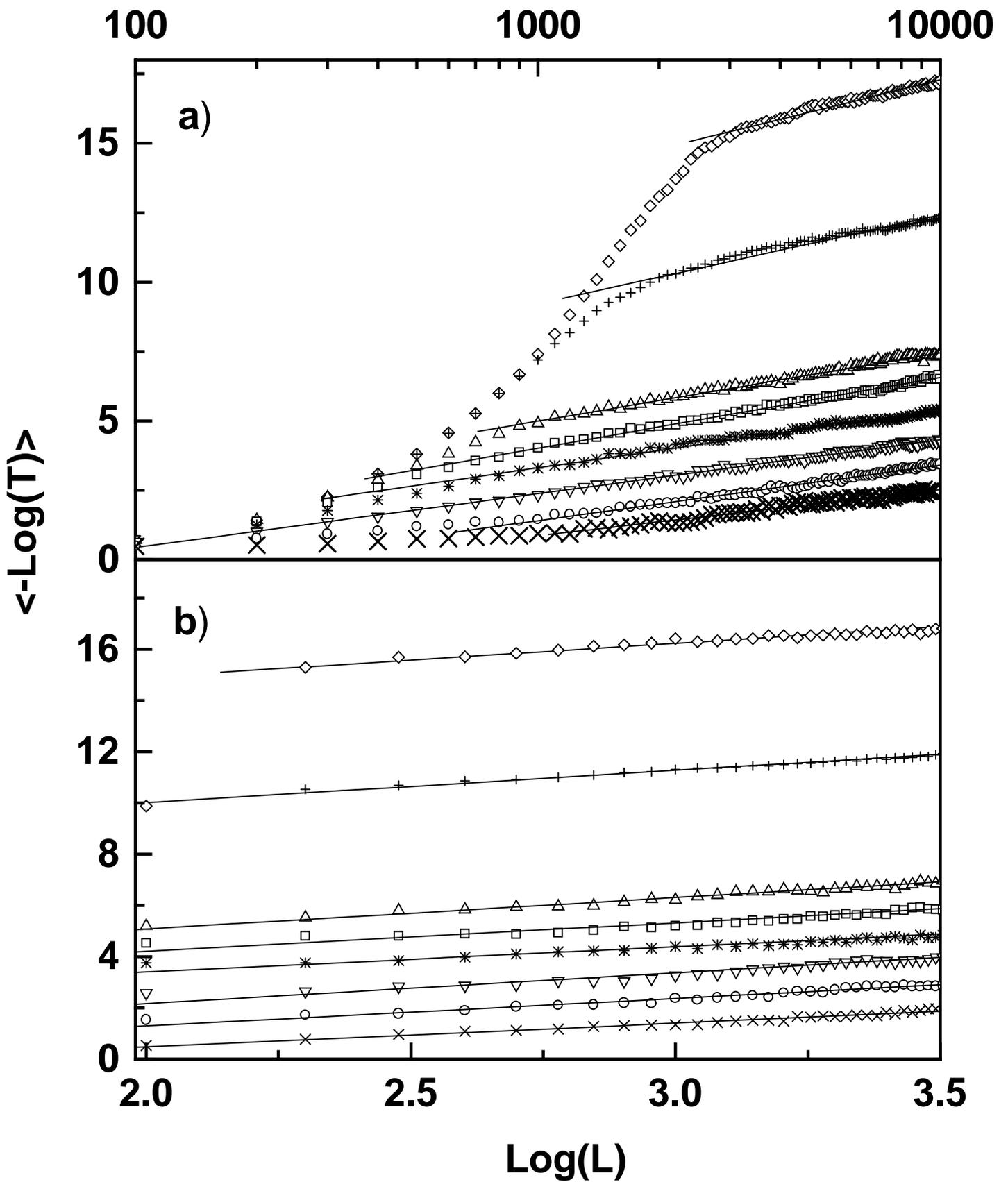
{\bf Fig.5 } $<-LogT>$ versus $LogL$ for $|\alpha|$ = $10^{-15}$ (open diamond), 
$10^{-10}$ (cross symbol +), $10^{-5}$ (open triangle up), 
$10^{-4}$ (open square), $10^{-3}$ (star symbol), $10^{-2}$ (open triangle down), 
$10^{-1}$ (open circle) and $1.$ (cross symbol x) for $\alpha < 0.$ , $W = 4.$, 
$E = 10$ and for $100$ disorder realizations a) Mixed case  b) Potential barriers.
Solid lines correspond to the power-law fittings. 

\bigskip

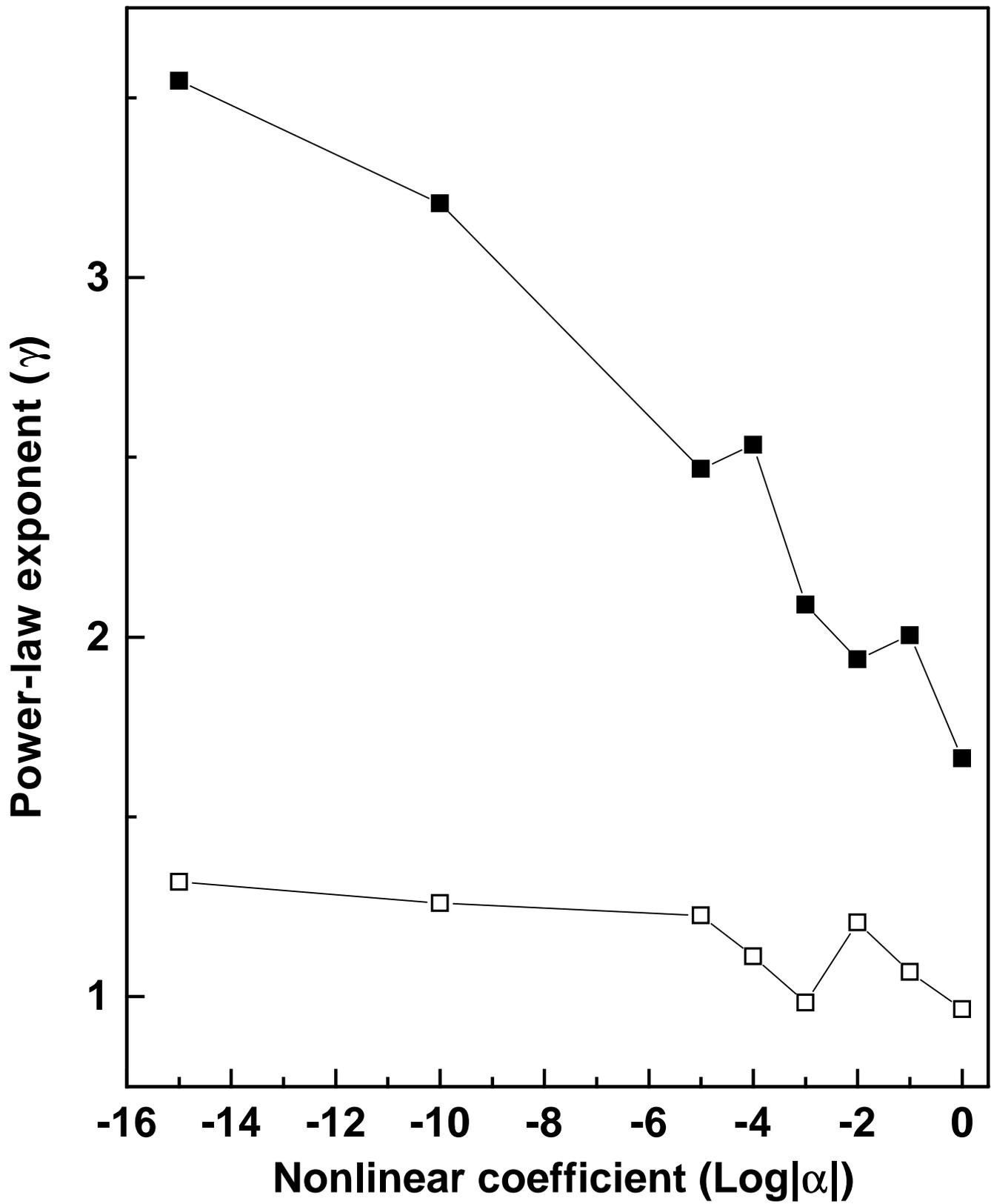
{\bf Fig.6 } The exponent $\gamma$ versus nonlinearity strength $Log(|\alpha|)$ 
for both mixed case (filled square) and potential barriers (open square). The solid 
lines are simply guides to the eye.
  
\end{document}